# Elucidation of microstructure of single-phase microcrystalline silicon based on crystallite size distributions

SANJAY K. RAM[1], MD. N. ISLAM[2] AND SATYENDRA KUMAR[3]
*Department of Physics, Indian Institute of Technology Kanpur, Kanpur-208016, India*

P. ROCA I CABARROCAS
*LPICM, UMR 7647 - CNRS - Ecole Polytechnique, 91128 Palaiseau Cedex, France*

**Abstract**. – Highly crystallized undoped hydrogenated microcrystalline silicon films prepared using $SiF_4$–$H_2$ mixture plasma were investigated at various stages of growth employing different microstructural probes. Our self-consistent results elucidate various aspects of the evolution of film microstructure, compositional changes and variations in crystallite size distributions with film growth. Inclusion of a bimodal crystallite size distribution in microstructural data analysis leads to results that are corroborative with those obtained from other microstructural tools, and yields a more physically accurate and coherent description of microcrystalline silicon film microstructure.

Silicon thin films have dominated the large area electronics applications such as solar cells and thin film transistors for active matrix addressing of displays [1,2,3,4,5]. The generic term "microcrystalline silicon" is used for a large variety of materials consisting of a crystalline phase composed of crystallite grains and their conglomerates, an amorphous or disordered phase and voids. The deposition of microcrystalline silicon ($\mu$c-Si:H) film involves the formation of an amorphous silicon (*a*-Si:H) phase, followed by a mixed phase (*a*-Si:H + $\mu$c-Si:H) material [1], and finally a fully crystallized single phase $\mu$c-Si:H material [6], in that order. The depositional conditions determine the layer thickness of each phase and the crystallinity of the material [1]. In plasma deposited $\mu$c-Si:H material, the processing history greatly influences the film microstructure, and there exists a marked inhomogeneity in microstructure and morphology at different stages of growth and in different phases of the material [7], which influences the optoelectronic properties [8] and the path of carrier transport [9,10].

An important aspect of such microstructural inhomogeneity in single phase $\mu$c-Si:H films is the evolution of the crystallites, a distribution in the sizes of crystallites, and their coalescence to form columns with film growth [11]. However, the crystallite size distributions (CSD) are not conventionally taken into consideration while analyzing microstructural data of $\mu$c-Si:H films, although it is known that the distribution of the crystallite size is an important parameter in the determination of the optoelectronic properties of the material. This is especially true in the single phase $\mu$c-Si:H material when the crystallinity is high and non-changing with film growth and there is no amorphous phase.

An accurate microstructural characterization of the material requires different approaches to modeling of the microstructural data, keeping in perspective the stage of film growth and the microstructural inhomogeneities associated with that particular stage of growth. A proper and physically accurate microstructural characterization alone can provide the necessary quantitative knowledge of the interrelationships between deposition parameters, film microstructure at different stages of growth, and optoelectronic properties, that would enable one to optimize the microstructure and the consequent optoelectronic properties in a specific and designed manner [8]. In this letter, we report on our study of highly crystallized $\mu$c-Si:H material with an aim to elucidate its microstructure at different stages of film growth with the help of modeling methods that yield quantitative information regarding the constituent phases throughout the growth stages. Our proposed methodology and approach to RS data analysis based on the inclusion of the CSD helps to form a composite picture of this material by bringing comparability to the analyses of the different characterization tools operating at different length scales.

For this study we have studied highly crystallized undoped $\mu$c-Si:H films deposited on corning 1737 substrates at a low substrate temperature in a parallel-plate glow discharge plasma enhanced chemical vapor deposition system operating at a standard rf frequency of 13.56 MHz, using high purity $SiF_4$, Ar and $H_2$ as feed gases. We carried out microstructural investigations on a large variety of such $\mu$c-Si:H films by applying a variety of characterization tools on the same samples: in-situ phase modulated spectroscopic ellipsometry (SE), bifacial Raman scattering, X-ray diffraction (XRD) and atomic force microscopy (AFM). Different microstructural series of a large number of samples were created by systematically varying gas flow ratios ($R$ = $SiF_4/H_2$) or substrate temperature ($T_s$ = 100–250˚C) for samples having different thicknesses ($d \approx$ 50-1200nm) [6,11]. In this letter, we present the results of samples representing different stages of film growth deposited under comparable conditions ($R$ = 1/1 and $T_s$ = 200°C).

[1] Corresponding author. E-mail: skram@iitk.ac.in , sanjayk.ram@gmail.com
[2] Present address: QAED-SRG, Space Application Centre (ISRO), Ahmedabad – 380015, India
[3] E-mail : satyen@iitk.ac.in

Figure 1 shows the imaginary part of the pseudo-dielectric function $<\varepsilon_2>$ spectra measured by SE on $\mu$c-Si:H samples at different growth stages: initial ($d$ = 62 nm), intermediate ($d$ = 450 nm) and final ($d$ = 920 nm). The broad shoulders in the SE spectrum near 3.4 eV ($E_1$) and 4.2 eV ($E_2$) are the signs of the presence of a microcrystalline material [12,13]. The associated changes in the bulk crystalline material with increase in film thickness is reflected in the increase in the intensity of the shoulder of $E_2$ peak, observed here for the thicker films ($d$ = 450 and 920 nm), indicating excellent crystallization. Experimental data was fitted using standard Bruggeman effective medium approximation [14]. The choice of the reference crystalline silicon (c-Si) is crucial in obtaining a good fit and reasonable description of the film structure. We have used published dielectric functions for low-pressure chemical vapor deposited polysilicon with large (pc-Si-l) and fine (pc-Si-f) grains [12,15].

For the modeling of SE results, we have considered the $\mu$c-Si:H film as a three-layered structure model: (i) A bottom layer interfacing with substrate (denoted here as BIL); (ii) A middle bulk layer (denoted as MBL); and (iii) A top surface roughness layer (denoted as TSL). However, in very thin films the bulk and interface layers are indistinguishable and are considered together as one layer (BL). Each of these layers has a variable composition of *a*-Si:H, microcrystallites (large and fine grains) and voids. The percentage volume fractions of the components are denoted in this study as: $F_{cl}$ (large grains), $F_{cf}$ (small grains), $F_a$ (*a*-Si:H) and $F_v$ (void content). The fitting results of SE measurements presented here have $\chi^2$ value <1 and lie well within 90% confidence limits. An equally important consideration was that the fitting should result in physically acceptable parameter values. Our choice of a layerwise structural model and use of large and fine grains as c-Si reference were found to be sufficient to explain the detailed microstructure of the film at a particular growth stage.

The fitting results of SE data of the films at an early stage of growth have revealed that the bulk of the material contains mainly small grains, an amorphous content and voids. For instance, the film of $d$ = 62 nm has the following composition: $F_{cf} \approx 73\%$, $F_a \approx 21\%$, $F_v \approx 6\%$. With film growth, a rise in the total crystalline fraction in the bulk of the material was seen, with an absence of amorphous content and an increase in the fraction of large grains, thus revealing a continuous rise in the $F_{cl}/F_{cf}$ ratio. In addition, the material becomes densely packed with the void fraction getting gradually and continuously reduced ($F_v \rightarrow 0$) with film growth. This trend is depicted in the composition of the bulk layer of the film with $d$ = 450 nm, where $F_{cl} \approx 37.6\%$ and $F_{cf} \approx 58.7\%$, and in the film with $d$ = 920 nm, where $F_{cl} \approx 45\%$ and $F_{cf} \approx 51\%$. A thin (20–30 nm) *a*-Si:H rich incubation layer with density deficit is found at the glass-film interface in most samples, except in those deposited at higher $R$ values (1/10 and 1/20), where a well-crystallized $\mu$c-Si:H layer is formed at the onset of deposition, and a separate incubation layer cannot be discerned in the fitting process [11].

In fig. 2 we have shown the 2D morphological images (2000nm × 2000 nm) of the film surface of the same samples belonging to three different stages of growth. The surfaces of films at initial stage of growth ($d \approx$ 62nm) is mostly ($\approx$ 90% of the surface area) covered with small grains of an average size $\approx$ 15–20 nm. Some of them coalesce to form a few larger $\approx$ 30–40 nm sized grains. As the film growth proceeds, reaching an intermediate stage of growth ($d \approx$ 450 nm), the surfaces of the films contain grains of average size of 75–90 nm; however, few of them are 150–175 nm in size. Conglomeration of grains is not observed up to this stage of film growth. At the final stages of film growth ($d \approx$ 920 nm), when the growth process achieves a steady state and the morphological reordering has reached a level of full maturity, the AFM image shows that the size of the conglomerate grains are 225–250 nm, which are composed of 150–200 nm sized grains. However, few small (70–80 nm sized) grains are dispersed in the surface adjoining the conglomerate grains. We note here that at every stage, two types of surface grains having different average sizes, one large sized and another small, are distributed in the film surface, though they cannot be casually compared to or be indicative of the sizes of crystallites in the bulk. In all the *R* (1/1 to 1/10) series, the XRD studies carried out on the samples have shown two different sizes of crystallites, having different crystallographic orientations. In this particular *R* = 1/1 series, smaller sized crystallites (≈ 9 nm) are seen to be oriented in the (111) direction, while crystallites having a larger size (average 23nm) show different orientations; the crystallites of the fully grown samples specifically demonstrate preferred orientation in the (400) direction.

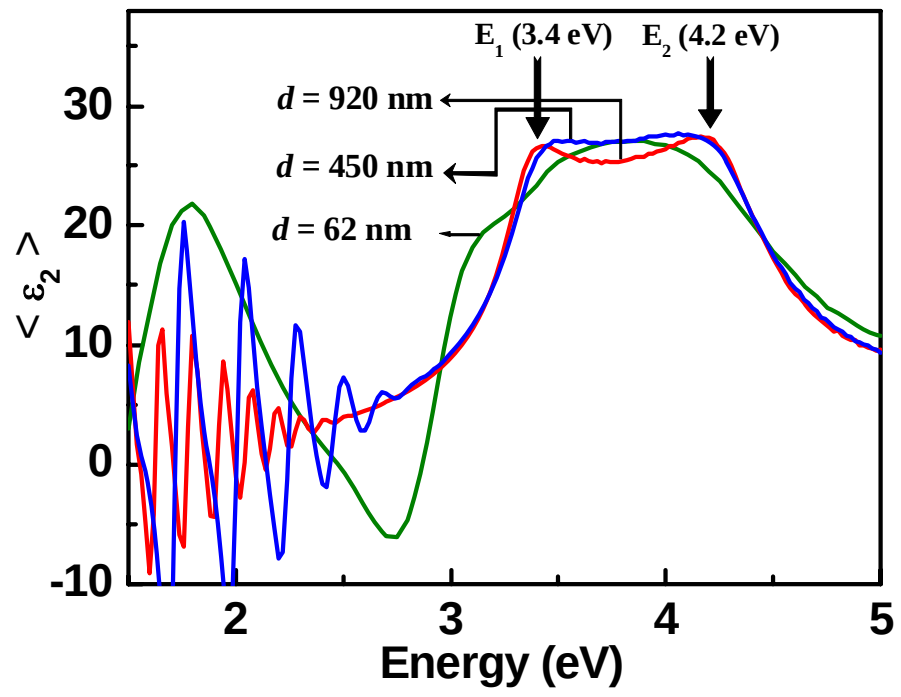

Fig. 1. Measured imaginary part of $<\varepsilon_2>$ spectra for $\mu$c-Si:H samples of different growth stages, $d$ = 62 nm, $d$ = 450 nm and $d$ = 920 nm.

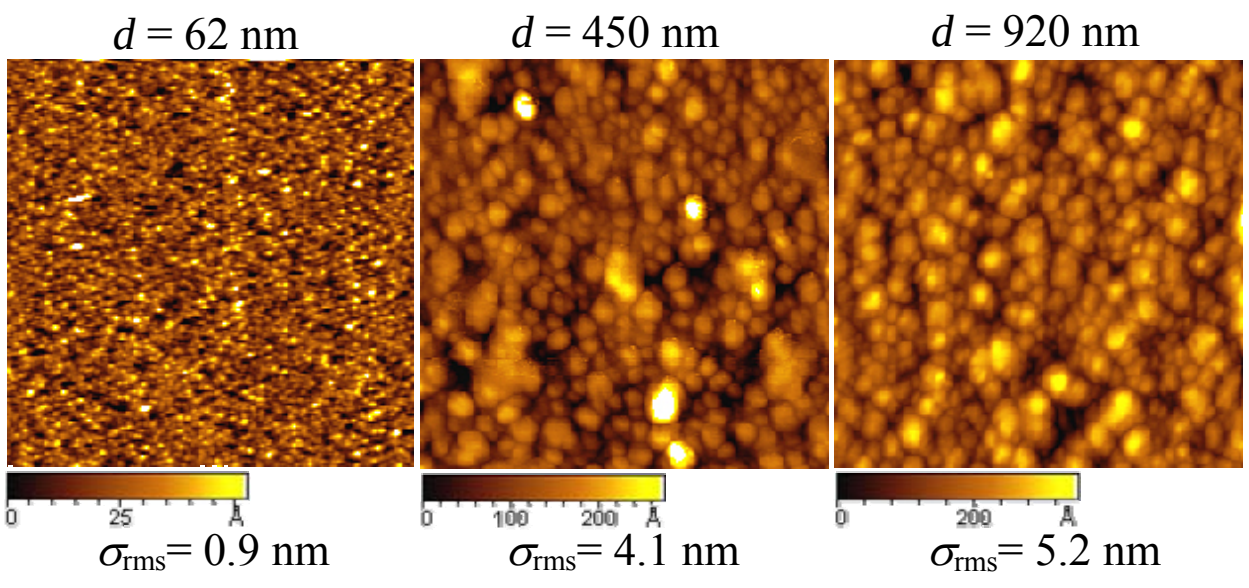

Fig. 2. 2D morphological images (2000 nm × 2000 nm) of the film surface of the same samples (fig. 1) belonging to three different stages of growth.

Now we come to the results of RS studies carried out on the same samples. Bifacial RS measurements were performed by focusing the light probe (He-Ne laser, λ=632.8 nm) from the top film surface [RS(F)], as well as from the bottom surface through the glass substrate [RS(G)]. Thus RS(G) provides information about the initial stages of film growth through film-substrate interface, while RS(F) depicts the terminal part of film growth, and when considered together, they reveal the inhomogeneities associated with the film growth. In fig. 3, we have shown the experimental data of bifacial RS of the same three films discussed above: RS(F) in figs. 3(a), (c) and (e); and RS(G) in figs. 3(b), (d) and (f).

A distribution in the sizes of crystallites is an established phenomenon in $\mu$c-Si:H material. However, the conventional approach to the deconvolution of RS data is done on the basis of an assumption of a single mean crystallite size and an amorphous phase, even in the single phase material. In the fully crystallized material, in the presence of a significant CSD, this may result in some inaccuracies in the estimation of mean crystallite size and the total crystalline volume fraction.

The SE, XRD and AFM studies of our samples have substantiated the presence of two different sizes of crystallites (or surface grains), although the numerical values of the sizes deduced from different tools are disparate, the reason for which is discussed later. The analysis of our SE results demonstrate that the large (pc-Si-l) and small (pc-Si-f) crystallites exist in significantly varying ratios for films at different growth stages and/or deposited under different $H_2$ dilution (effect of $H_2$ dilution is being reported elsewhere). These significant variations in both the sizes and volume fractions of crystallites in our $\mu$c-Si:H material are prohibitive to the generalized conventional approach to Raman data analysis assuming a single mean crystallite size [16,17,18] if a realistic and unambiguous microstructural picture of the material is desired. Therefore, a suitable deconvolution model applicable to the RS profiles of our samples would need to be accurate and physically rational, appropriate to our material and at the same time, inclusive of the input from the SE studies. In our work, we have used the model described in Ref. [19,20] for obtaining the CSD in Si nanostructures. According to this model, $\Phi(L)$ representing the CSD of an ensemble of spherical crystallites, total Raman intensity profile for the whole ensemble of nanocrystallites becomes:

$$I(\omega, L_0, \sigma) = \int \Phi(L) I'(\omega, L)\, dL \qquad (1)$$

For a normal CSD, $\Phi(L)$ is given as:

$$\Phi(L) = \frac{1}{\sqrt{2\pi\sigma^2}} \exp\left[-\frac{(L-L_0)^2}{2\sigma^2}\right] \qquad (2)$$

where the mean crystallite size $L_0$ and the standard deviation $\sigma$ are the characteristics of the CSD. By putting Eq.(2) into Eq.(1) and then integrating the results over the crystallite sizes *L*, and by restricting the dispersion parameter $\sigma$ to be less than $L_0/3$ one gets the modified Raman intensity profile as:

$$I(\omega, L_0, \sigma) \propto \frac{f(q) q^2 \exp\left\{-\dfrac{q^2 L_0^2 f^2(q)}{2\alpha}\right\}}{\{\omega - \omega(q)\}^2 + (\Gamma_0/2)^2}, \qquad (3)$$

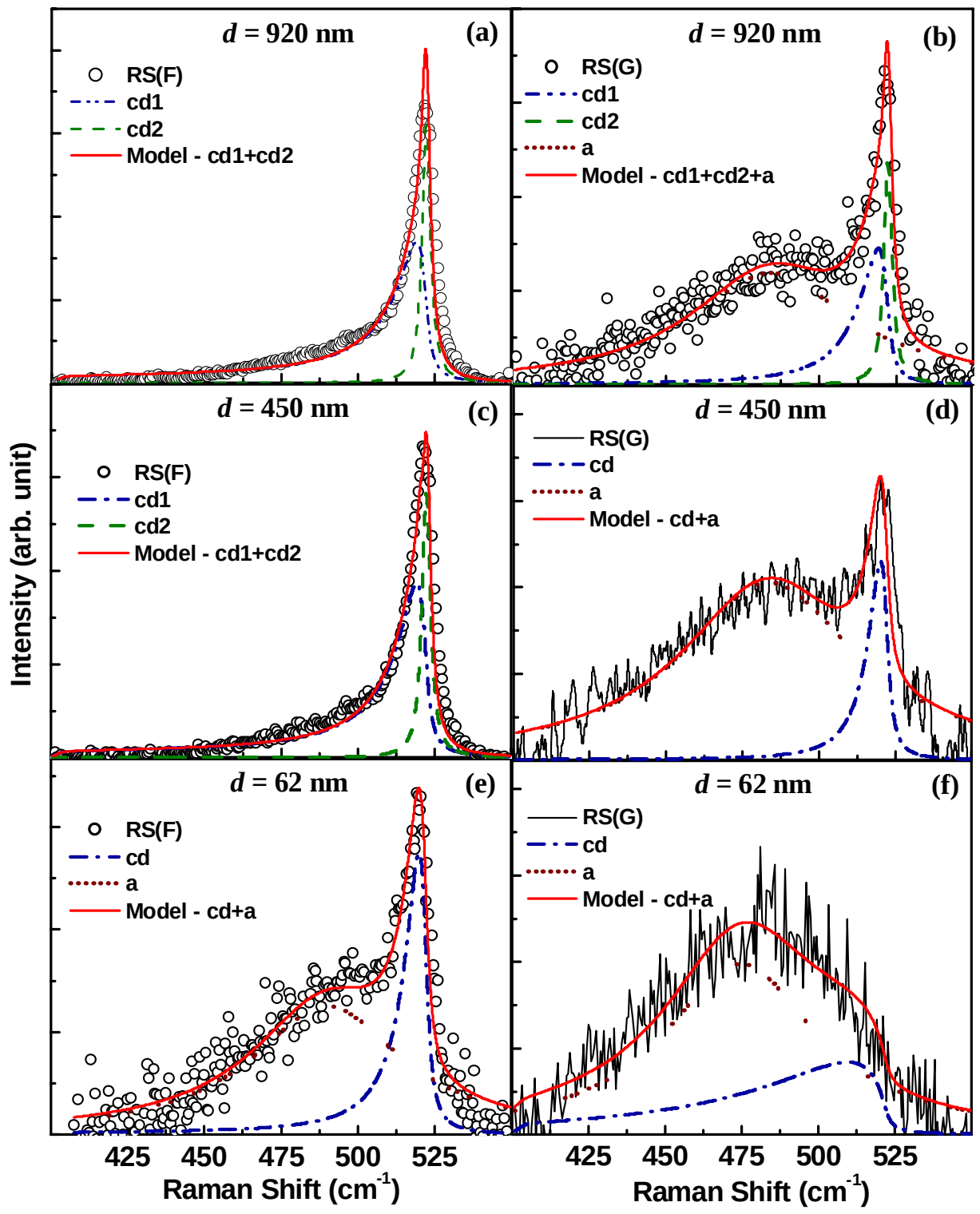

Fig. 3. Deconvolution of RS profiles of same samples (fig. 1) of different *d*. The parts (a), (c) and (e) depict RS(F) profiles; (b), (d) and (f) show RS(G) profiles. The respective deconvolution details and fitting models are mentioned in each part of the figure.

where the parameter $f(q) = 1\Big/\sqrt{\left(1+\frac{q^2\sigma^2}{\alpha}\right)}$, which incorporates the distribution broadening parameter $\sigma$ into the Raman intensity profile. The Gaussian distribution is found to explain our experimental results adequately; however, other distributions such as log normal may also be employed without any loss of generality in our approach [19,20].

It is evident from the fig. 3 that except in the thin sample ($d$ = 62 nm), an amorphous hump (a broad peak at 480 cm$^{-1}$) is absent in the RS(F) profiles of our samples. An asymmetry in the Raman lineshape of RS(F) of the other thick samples, seen as a low energy tail, arises from the distribution of smaller sized crystallites. An explicit amorphous hump is present in the RS(G) profiles of all the samples, as evident from figs. 3(b), (d) and (f).

Applying such considerations as discussed above basically gives rise to three different models which can be applied to the deconvolution of the RS profiles of our samples:

(i) Model "cd1+cd2+a" incorporating a bimodal CSD [small (cd1) and large (cd2) crystallites] along with an amorphous phase;
(ii) Model "cd1+cd2" where a bimodal CSD without an amorphous background is considered; and
(iii) Model "cd+a" having only one type of CSD (mainly the contribution from small sized crystallites) with an amorphous background.

These fitting models were applied to all the RS profiles of the samples. While the goodness of fit is usually identifiable visually and by statistical analysis, some closer fits required that to choose the most appropriate fit, we assess its physical plausibility, the corresponding SE data and the direction of RS measurement. The direction of RS measurement is important because the Raman collection depth for $\mu$c-Si:H material at 632.8 nm (He-Ne red laser) is ≈ 500 nm, which has to be taken into account along with the thickness of the sample to understand the data collected by RS from each direction. The final deconvolution results consisting of the total fit, and the deconvoluted peaks are shown in the same fig. 3 for each sample. It can be seen in these graphs that consideration of an amorphous content in the samples at the early stages of growth and a bimodal CSD in samples that have attained a certain stage of film growth, are rational approaches that provide a valid and accurate fitting of the RS data.

The deconvolution of RS profiles of our $\mu$c-Si:H samples using a size distribution of large (≈ 70–80 nm) and small (≈ 6–7 nm) grains bears out the bimodal CSD recognized by the different microstructural probes. However, the sizes obtained by various characterization techniques may show some differences that can be understood as follows. RS is an optical measurement technique, which elicits only the average sizes of the large and small grains, irrespective of their orientations. The deconvolution models are quite insensitive for large crystallites. In contrast, XRD measures the average crystallite size having a particular orientation, and is influenced by the averaging of the sizes of all the large and small grains present in a certain orientation. A cursory look at the AFM results suggests a rather marked deviation from the RS values, especially in the fully grown samples, in regards to the sizes of small and large grains. This is because AFM has a limitation of resolution resulting from the large conglomerates present at this stage of film growth.

Thus, the analyses of SE, bifacial RS and other microstructural probes are strongly corroborative and demonstrate the importance of differential treatment while modeling microstructural data of $\mu$c-Si:H films, according to the stage of growth. For a better visual understanding, the results obtained from various characterization probes are compositely summarized in a schematic illustration in fig. 4 for the samples representing different growth stages. Left column of the figure shows quantitative microstructural information obtained from SE analyses of the different film layers, while the right column shows results from AFM and RS(F) and RS(G) analyses. The thickness of the top surface layer as deduced by SE is substantiated by the roughness ($\sigma_{rms}$) elicited by AFM measurements and the existence of a linear correlation between the two has been shown in our previous report [11]. There is good qualitative agreement between the RS and SE data in terms of the microstructural growth trend, especially the fractional compositions of the constituent grains having different mean sizes. A limited quantitative match is expected between the results of different tools due to the different length scales of operation and limitations inherent to the principle of operation of the tool. The penetration depth of Raman is indicated by arrows in fig. 4 to emphasize the differences in the data collected by SE and RS for the same samples.

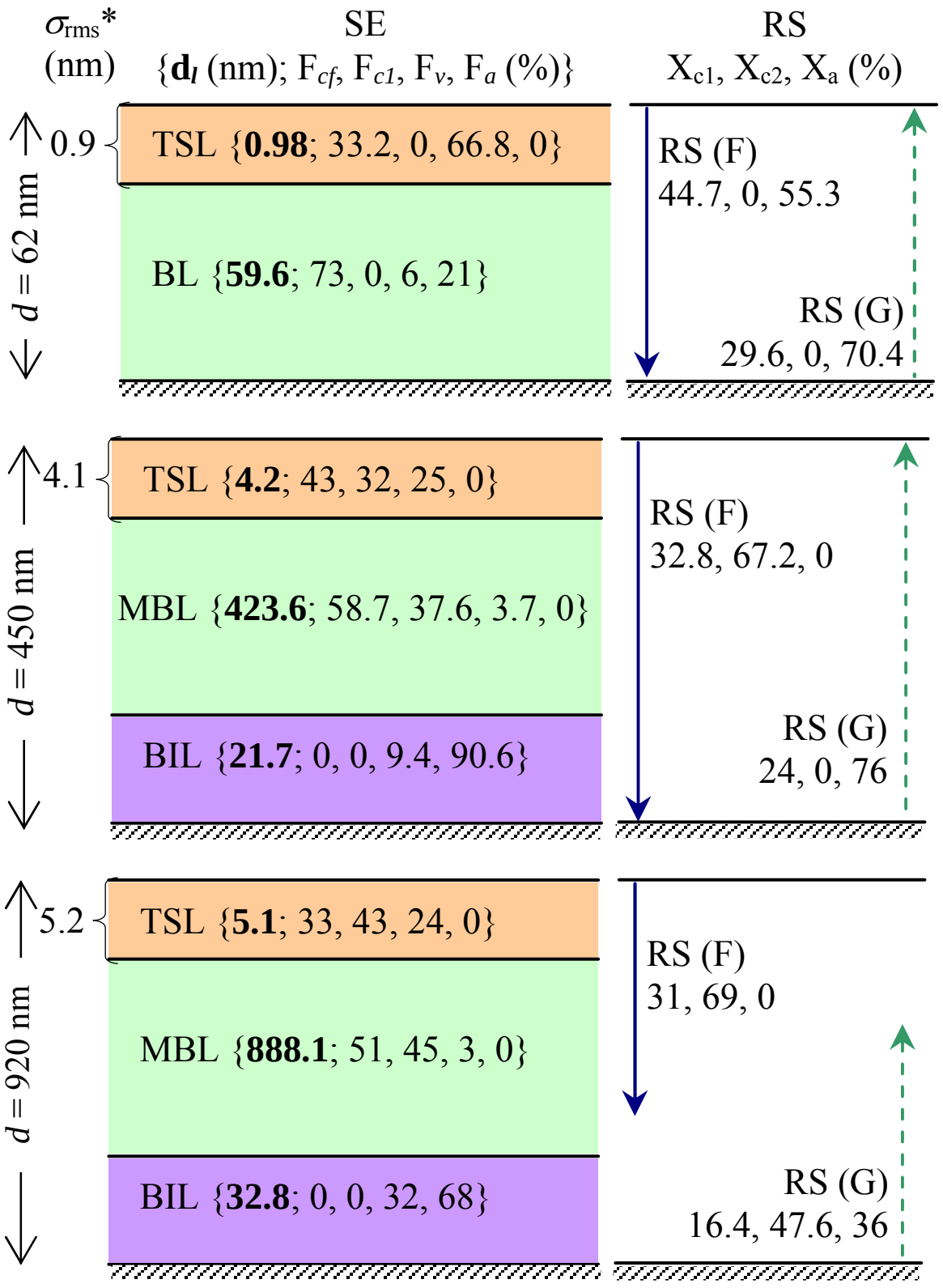

Fig. 4. A composite schematic illustration of the analyses of microstructural data obtained from various probes for samples of different $d$. The left column shows the SE analysis, while the right column shows bifacial RS data. $\sigma_{rms}$* denotes root mean square value of surface roughness obtained from AFM. $\mathbf{d}_l$ denotes thickness of the particular layer. Percentage fractions deduced from RS are: $X_{c1}$ (small crystallites), $X_{c2}$ (large crystallites) and $X_a$ (amorphous content). The arrows in the RS column represent the penetration depth (≈ 500 nm) of Raman excitation beam. ($R$ = 1/1, $T_s$=200 °C for all samples).

The presence of two distinct and remarkably different mean sizes of crystallites has far reaching implications for the optoelectronic properties of the material. The overall morphological evolution of the crystallite grains and their conglomerate columns can be well correlated to the changing fractions of constituent large and small grains (large grains specifically), which is especially significant from an electrical transport point of view in this type of single-phase material. This has been explored in our work on the optoelectronic properties of this material, which has been reported elsewhere [8,21].

In conclusion, our study of highly crystallized $\mu$c-Si:H films using different microstructural probes elucidates a self-consistent picture of microstructural evolution with film growth. The quantitative information derived from all the studies indicates a change in the volume fractions of different sized constituent grains taking place with film growth. The corroborative results obtained from different techniques make out a strong case for a RS data modeling approach based on the inclusion of a bimodal crystallite size distribution for the analysis of microstructural data of $\mu$c-Si:H films.

---